\documentclass[conference, 10pt]{IEEEtran}

\usepackage{cite}
\usepackage[cmex10]{amsmath}
\usepackage{url}
\usepackage{graphicx}
\usepackage{color}
\usepackage{placeins}
\usepackage{float}
\usepackage{hyperref}
\usepackage{tabularx,colortbl}
\usepackage{ifthen}

\makeatletter

\newcounter{author}
\renewcommand{\author}[2][]{
   \stepcounter{author}
   \@namedef{author@\theauthor}{#2}
   \@namedef{authorlabel@\theauthor}{#1}
}

\newcounter{address}
\newcommand{\address}[2][]{
   \stepcounter{address}
   \@namedef{address@\theaddress}{#2}
   \@namedef{addresslabel@\theaddress}{#1}
}

\newcommand{\alsep}{and}

\def\newmaketitle{\par%
  \begingroup%
  \normalfont%
  \def\thefootnote{}
  \def\footnotemark{}
  \let\@makefnmark\relax
  \footnotesize
  \footnotesep 0.7\baselineskip
  \normalsize%
  \twocolumn[\thenewmaketitle\@IEEEaftertitletext]%
  \if@IEEEusingpubid
     \enlargethispage{-\@IEEEpubidpullup}%
  \fi
  \endgroup
  \setcounter{footnote}{0}\let\maketitle\relax\let\@maketitle\relax
  \gdef\@thanks{}%
  \let\thanks\relax}

\def\thenewmaketitle{
  \newpage
  \begin{center}%
    \vskip0.2em{\Huge\@IEEEcompsoconly{\sffamily}\@IEEEcompsocconfonly{\normalfont\normalsize\vskip 2\@IEEEnormalsizeunitybaselineskip
   \bfseries\large}\@title\par}\vskip1.0em\par%
    \vspace{1ex}
    \newcounter{c@author}
    \newcounter{c@tmp}
    \ifthenelse{\value{author}=2}{%
      \newcommand{\liand}{ and }}{%
      \newcommand{\liand}{, and }}
    \ifthenelse{\value{address}<2}{%
      \@nameuse{author@1}%
      \stepcounter{c@author}%
      \whiledo{\value{c@author}<\value{author}}{%
        \setcounter{c@tmp}{\value{author}}%
        \addtocounter{c@tmp}{-\value{c@author}}%
        \ifthenelse{\value{c@tmp}=1}{%
          \renewcommand{\alsep}{\liand}}{\renewcommand{\alsep}{, }}%
        \stepcounter{c@author}\alsep \@nameuse{author@\thec@author}}\\%
    }
    {
      \@nameuse{author@1}${}^{(\ref{\@nameuse{authorlabel@1}})}$%
      \stepcounter{c@author}%
      \whiledo{\value{c@author}<\value{author}}{%
      \setcounter{c@tmp}{\value{author}}%
      \addtocounter{c@tmp}{-\value{c@author}}%
      \ifthenelse{\value{c@tmp}=1}{%
        \renewcommand{\alsep}{\liand}}{\renewcommand{\alsep}{, }}%
      \stepcounter{c@author}\alsep \@nameuse{author@\thec@author}%
        ${}^{(\ref{\@nameuse{authorlabel@\thec@author}})}$%
      }
    }
    \vspace{0.2ex}

    \ifthenelse{\value{address}>0}{%
      \ifthenelse{\value{address}=1}{
        {\@nameuse{address@1}}
      }
      {
        \newcounter{c@address}

        \begin{center}
        \whiledo{\value{c@address}<\value{address}}
        {
          \refstepcounter{c@address}
            ${}^{(\thec@address)}$\,%
              \label{\@nameuse{addresslabel@\thec@address}}%
              \@nameuse{address@\thec@address}\\ %
        }
        \end{center}
      } 
    }
    {
      \relax
    }
  \end{center}
}

\makeatother

\title{AAFIYA: Antenna Analysis in Frequency-domain for Impedance and Yield Assessment}

\author[org1]{M.~F.~H.~Seikh}
\author[org2]{R.~Jarvis}
\author[org2]{J.~Stiles}

\address[org1]{Department of Physics and Astronomy, University of Kansas, Lawrence, KS 66045 USA}
\address[org2]{Department of Electrical Engineering \& Computer Science, University of Kansas, Lawrence, KS 66045 USA}

\begin{document}

\newmaketitle

\begin{abstract}
This paper presents AAFIYA (Antenna Analysis in Frequency-domain for Impedance and Yield Assessment), a modular Python toolkit for automated characterization of radio-frequency antennas using measurement and simulation data. The toolkit provides a unified workflow for processing S-parameters, impedance, realized gain, beam patterns, polarization metrics, and calibration-based yield estimation, with support for standard Touchstone files and beam pattern data. AAFIYA is validated using measurements from an electromagnetic anechoic chamber involving Log Periodic Dipole Array (LPDA) reference antennas and Askaryan Radio Array (ARA) Bottom Vertically Polarized antennas over 100–850 MHz. Extracted metrics, including impedance matching, realized gain patterns, vector effective lengths, and cross-polarization ratio, are compared against full-wave simulations from HFSS and WIPL-D, showing good agreement across frequency and angle. The results demonstrate that AAFIYA enables accurate, reproducible, and publication-ready antenna analysis, and provides a flexible foundation for future extensions, including automated optimization and data-driven antenna design.
\end{abstract}


\section{Introduction}
The rapid development and deployment of radio-frequency (RF) and microwave systems in scientific, industrial, and commercial applications has increased the demand for rigorous, reproducible, and transparent antenna characterization workflows. Accurate analysis of antenna parameters such as reflection ($S_{11}$), transmission ($S_{21}$), impedance, voltage standing wave ratio (VSWR), and beam patterns is fundamental for optimizing antenna design, evaluating system performance, and ensuring compatibility with experimental requirements~\cite{pozar2011microwave, balanis2016antenna}.  

While a variety of commercial and academic tools exist for antenna analysis, researchers often face limitations related to data accessibility, reproducibility, customization, and publication-quality visualization. Measurement and simulation outputs are commonly stored in heterogeneous formats, including Touchstone ($.s1p$, $.s2p$) files and beam pattern text ($.txt$) files, motivating the need for flexible and transparent analysis pipelines.

AAFIYA (\textbf{A}ntenna \textbf{A}nalysis in \textbf{F}requency-domain for \textbf{I}mpedance and \textbf{Y}ield \textbf{A}ssessment) is a Python-based toolkit developed to address these challenges by providing a comprehensive and extensible platform for antenna characterization.\footnote{The AAFIYA source code and documentation are publicly available at \url{https://github.com/Mohammad-Neutrino/AAFIYA}.} Designed for both research and pedagogical applications, AAFIYA ingests measured or simulated antenna data files (e.g., S-parameter and beam pattern files), computes user selected antenna performance metrics, and produces reproducible, publication-quality visualizations and summary.

The toolkit supports the full antenna characterization workflow, including analysis of $S$-parameters, impedance, VSWR, group delay, Smith chart representations, realized gain, beam patterns, polarization metrics, and calibration-based yield estimation. AAFIYA emphasizes modular design, consistent units, and publication-quality plotting, allowing it to function both as a standalone analysis tool and as a component within larger experimental or simulation workflows.

In this paper, we describe the design and application of AAFIYA and demonstrate its performance using representative antenna measurements. Results from Log Periodic Dipole Array (LPDA) reference antennas and Askaryan Radio Array (ARA) Bottom Vertically Polarized (BVPol) antennas~\cite{Seikh:2024_EPJST} are presented and validated against full-wave electromagnetic simulations, illustrating the accuracy and robustness of the analysis pipeline.


\section{Toolkit Overview}

AAFIYA is implemented as a modular Python toolkit that separates antenna data ingestion, metric computation, and visualization into reusable components. This design supports both batch and interactive workflows, making the toolkit suitable for experimental campaigns, simulation post-processing, and rapid prototyping.

The project is organized into two main components: core source modules and standalone analysis scripts. The core functionality resides in the \texttt{src/} directory. The \texttt{antenna\_io.py} module handles data ingestion from standard Touchstone S-parameter files and beam pattern text files, providing consistent access to frequency-dependent complex data. Antenna performance metrics are implemented in \texttt{antenna\_metrics.py}, including reflection and transmission coefficients, VSWR, complex impedance, group delay, realized gain, beamwidth, front-to-back ratio (F/B), realized vector effective length (RVEL), and cross-polarization ratio (XPR). These quantities are computed using standard definitions and calibration relationships widely used in antenna theory and RF measurements \cite{pozar2011microwave,balanis2016antenna}, with all calculations implemented in a vectorized and unit-consistent manner. Plotting and visualization routines are centralized in \texttt{antenna\_plot.py} to ensure consistent formatting and reproducibility across analyses. Reproducibility is ensured through consistent plotting conventions, including fixed axis definitions, units, labeling, color mapping, and line styles, allowing figures generated from different datasets or experiments to be directly compared without manual reformatting.

Standalone scripts located in the \texttt{scripts/} directory perform specific analysis and visualization tasks. These include scripts for S-parameter and impedance analysis, realized gain extraction, two- and three-dimensional beam pattern visualization, frequency-dependent beamwidth and front-to-back ratio studies, and polarization analysis. Additional scripts enable direct comparison between measured antenna data and electromagnetic simulations, overlaying measured $S_{11}$ and realized gain patterns with results obtained from HFSS and WIPL-D simulations within the NuRadioMC framework~\cite{NuRadioMC}.

This separation of concerns allows users to readily adapt AAFIYA to new datasets or experimental configurations while maintaining a transparent and reproducible analysis workflow. The modular structure also facilitates extension of the toolkit to additional antenna metrics or integration with larger simulation and data analysis pipelines.


\section{Measurement and Calibration}

All measurements processed with AAFIYA were performed in an electromagnetic anechoic chamber using a vector network analyzer (VNA)–based setup, with a fixed antenna separation of 6.5 m between the reference antenna and the antenna under test (AUT). Both S-parameter and beam pattern measurements were conducted to characterize reference and AUT configurations across a wide frequency range.


\subsection{S-Parameter Measurements}

For two-port characterization, two LPDA antennas were arranged in a face-to-face configuration and connected to a VNA, enabling measurement of the full $2 \times 2$ S-parameter matrix ($S_{11}$, $S_{21}$, $S_{12}$, $S_{22}$) over the frequency range 100--850~MHz. In a second measurement set, the receive-side LPDA was replaced with an ARA BVPol antenna, which was characterized using the same procedure. This approach provides reference data from a well-characterized antenna as well as measurements for the AUT under identical conditions.


\subsection{Beam Pattern Measurements}

Beam pattern measurements were performed using the LPDA as a transmitting reference antenna positioned at a fixed location in the anechoic chamber. The AUT was mounted on a motorized turntable at a separation of 6.5~m and rotated through $360^\circ$ in $1^\circ$ increments. Measurements were repeated across the full frequency sweep for multiple principal planes (V-Plane, H-Plane, and Z-Plane) and for both co-polarized and cross-polarized configurations. Equivalent scans using LPDAs on both sides were performed to provide reference measurements under identical geometric and instrumental conditions.


\subsection{Gain Extraction and Calibration}

AAFIYA implements a calibration-based approach to extract the realized gain pattern of the AUT using measurements from a well-characterized reference antenna. First, the transmission coefficient $S_{21,\mathrm{ref}}(0^\circ,f)$ is measured at boresight between two identical reference antennas. The boresight gain of the reference antenna, $G_{\mathrm{ref}}(0^\circ,f)$, is obtained from electromagnetic simulations, manufacturer data, or a dedicated calibration measurement using the Friis transmission equation,
\begin{equation}
G_{\mathrm{ref}}(0^\circ,f) = \frac{|S_{21,\mathrm{ref}}(0^\circ,f)| \cdot 4\pi R}{\lambda},
\label{eq:friis_ref}
\end{equation}
where $R$ is the antenna separation and $\lambda$ is the wavelength. In this work, the reference antenna gain was extracted directly from measurements using two identical LPDA antennas arranged in a face-to-face configuration and calibrated via \eqref{eq:friis_ref}.

After calibration, the AUT replaces one reference antenna and the transmission coefficient $|S_{21,\mathrm{AUT}}(\theta,f)|$ is measured as a function of angle and frequency. The realized gain of the AUT is then computed as
\begin{equation}
G_{\mathrm{AUT}}(\theta,f) =
\frac{|S_{21,\mathrm{AUT}}(\theta,f)|}{|S_{21,\mathrm{ref}}(0^\circ,f)|}
\cdot G_{\mathrm{ref}}(0^\circ,f).
\end{equation}
Known system losses and power offsets are accounted for during this calculation. This procedure enables traceable and reproducible extraction of realized gain patterns for comparison with simulation and further analysis.


\section{Results}

This section presents representative results obtained using AAFIYA for antenna characterization, including S-parameter analysis, realized gain extraction, and validation against full-wave electromagnetic simulations. Measurements are shown for both LPDA reference antennas and ARA BVPol antennas.

\subsection{S-Parameter Analysis}

\begin{figure}[t]
    \centering
    \includegraphics[width=0.9\columnwidth]{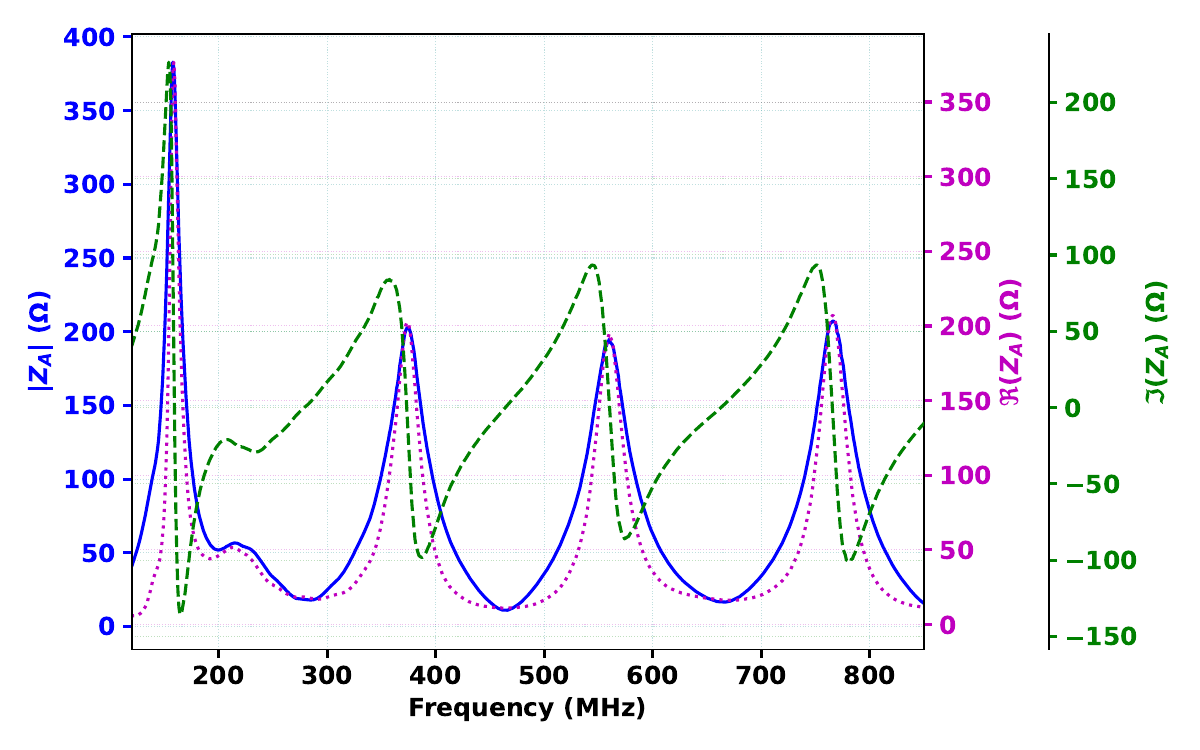}
    \includegraphics[width=0.9\columnwidth]{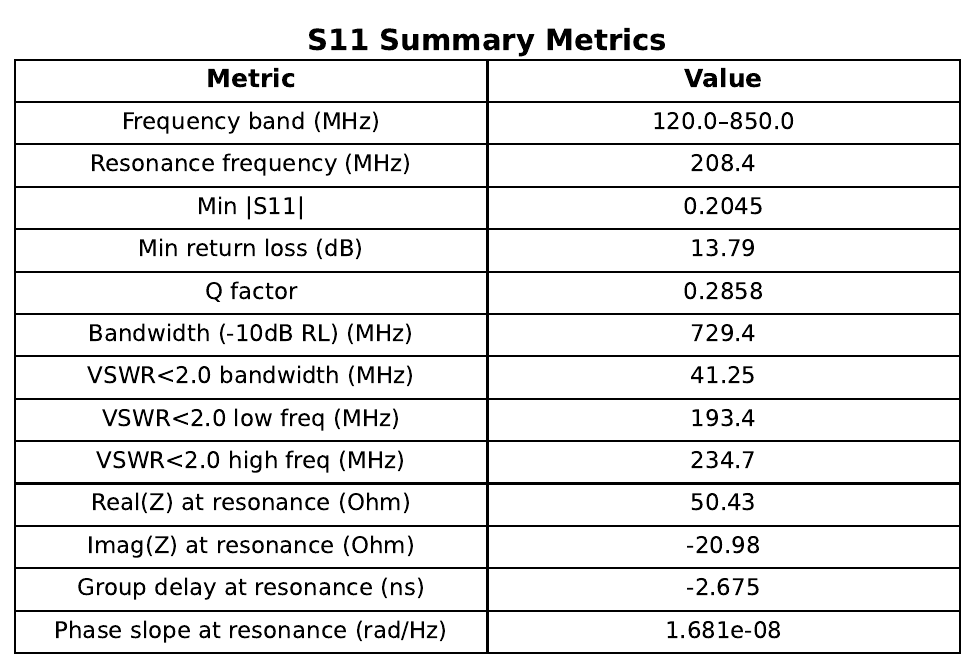}
    \caption{ARA BVPol S-parameter characterization. \textbf{Top:} Measured complex input impedance as a function of frequency. \textbf{Bottom:} Summary of $S_{11}$-derived metrics, including resonance frequency, return loss, bandwidth, VSWR performance, and impedance at resonance.}
    \label{fig:ara_bvpol_s11}
\end{figure}

S-parameter measurements provide insight into impedance matching, transmission behavior, and port isolation of the measured antenna systems. Fig.~\ref{fig:ara_bvpol_s11} shows the measured complex input impedance and a summary of key $S_{11}$-derived metrics for the ARA BVPol antenna. The results indicate broadband behavior with a clear resonance near the design frequency and good impedance matching across a wide operational band.

\begin{figure}[t]
    \centering    \includegraphics[width=0.9\columnwidth]{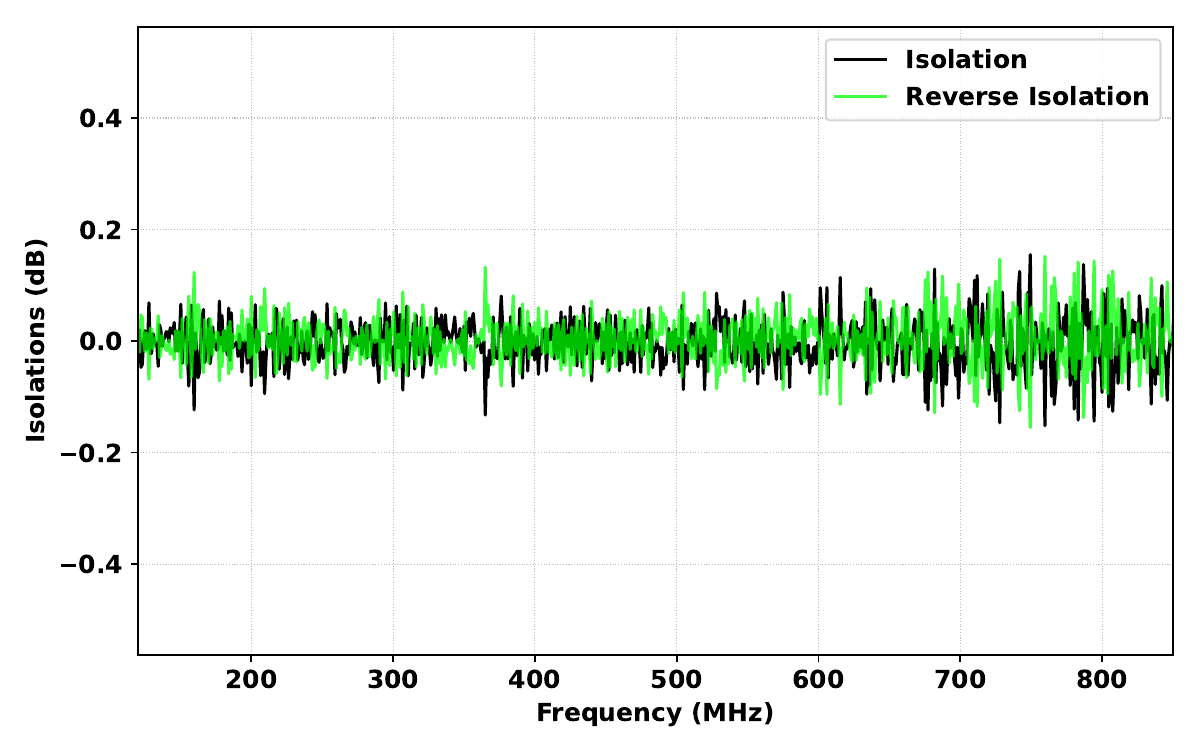}
    \caption{Isolation and reverse isolation as a function of frequency for the Tx–Rx LPDA H-Plane configuration, demonstrating strong port separation across the operational band.}
    \label{fig:lpda_isolation}
\end{figure}

For the reference Tx–Rx LPDA configuration, isolation and reverse isolation measurements are shown in Fig.~\ref{fig:lpda_isolation}. Strong port isolation is observed across the measured frequency range, confirming minimal crosstalk and validating the use of the LPDA as a reference antenna for gain calibration.

\subsection{Gain Pattern Analysis}

\begin{figure}[t]
    \centering
    \includegraphics[width=0.75\columnwidth]{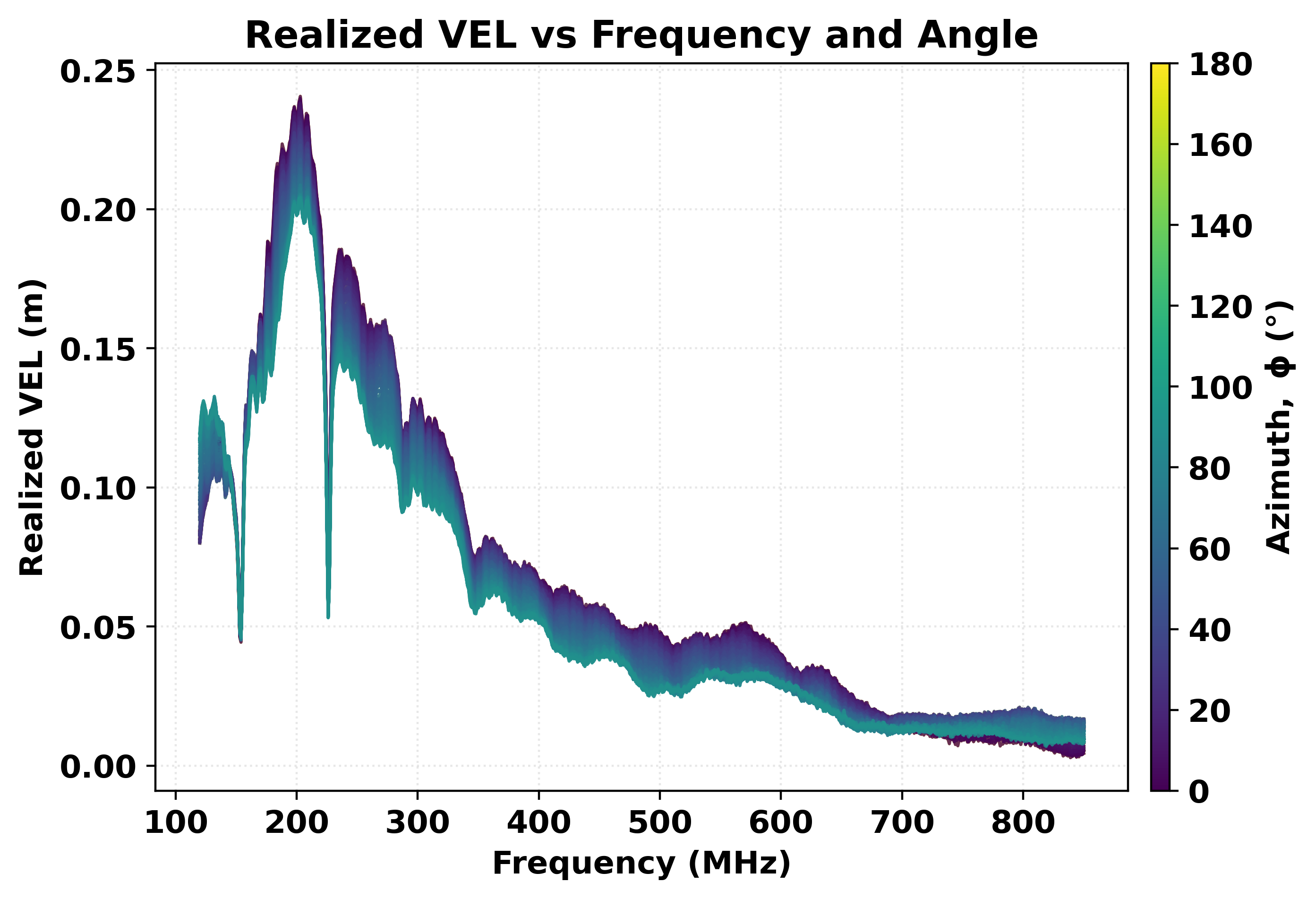}
    \includegraphics[width=0.75\columnwidth]{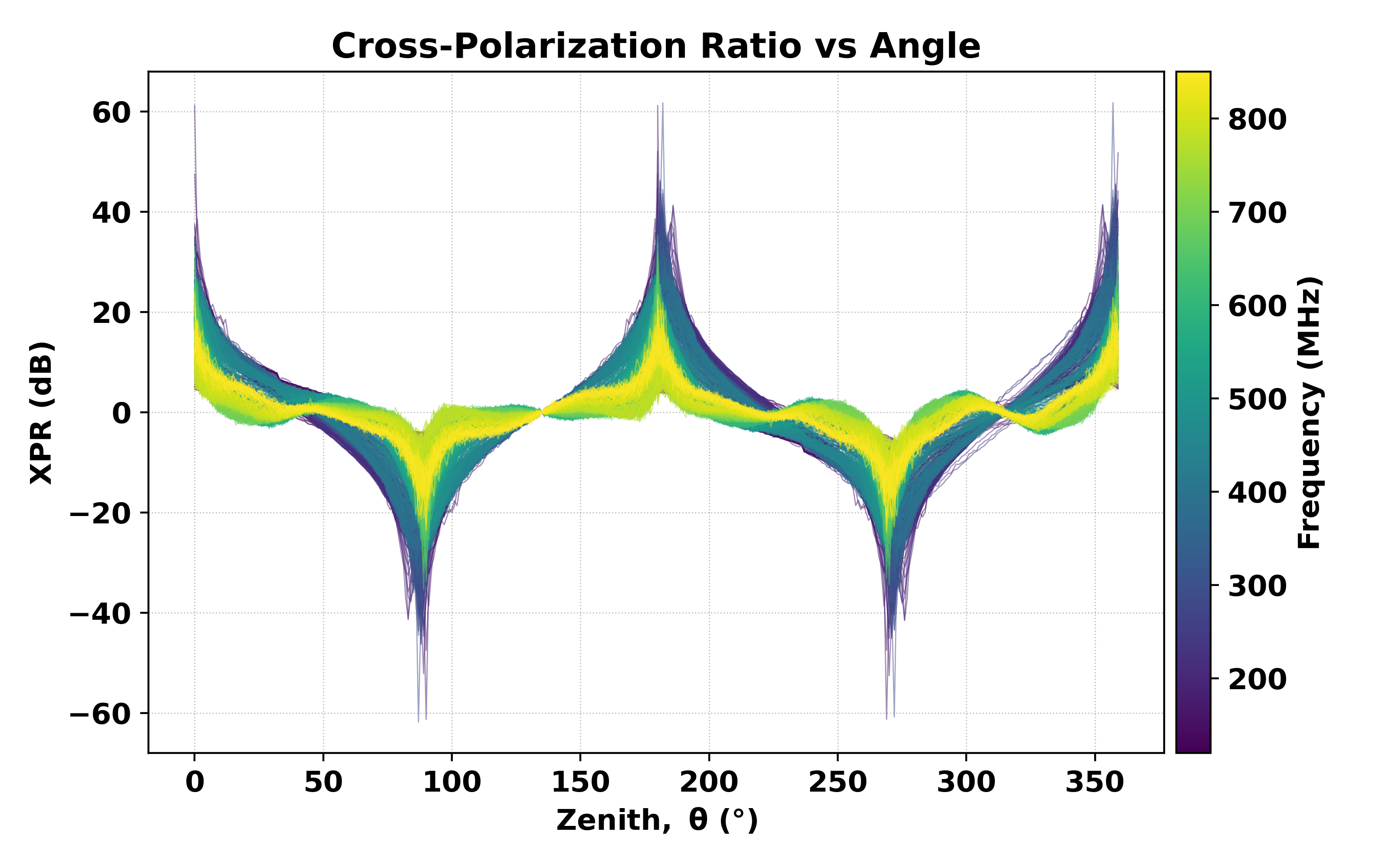}
    \includegraphics[width=0.7\columnwidth]{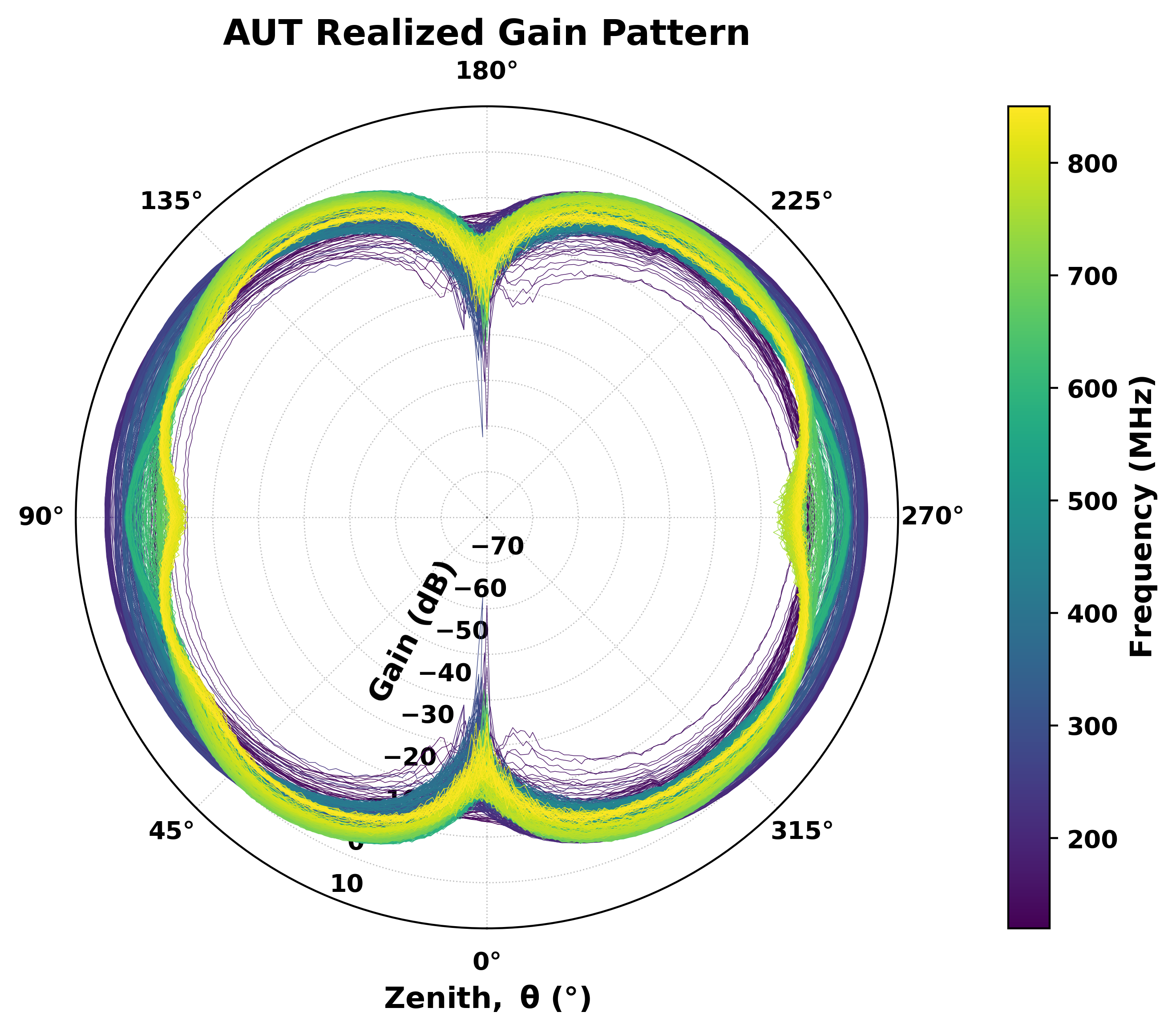}
    \caption{ARA BVPol gain characteristics. \textbf{Top:} RVEL as a function of frequency and azimuth. \textbf{Middle:} Cross-polarization ratio (XPR) in the zenith plane, quantifying polarization purity. \textbf{Bottom:} Realized gain pattern in the zenith plane.}
    \label{fig:ara_bvpol_gain}
\end{figure}

Realized gain characterization was performed for both the ARA BVPol and LPDA antennas using the calibration procedure described in Section~III. Fig.~\ref{fig:ara_bvpol_gain} summarizes the principal gain-related results for the ARA BVPol antenna. The realized vector effective length (RVEL) highlights the frequency-dependent sensitivity and angular response, while the zenith-plane realized gain pattern reveals the main lobe structure and angular coverage. Cross-polarization ratio (XPR) measurements indicate good polarization purity across much of the operational bandwidth.

\begin{figure}[t]
    \centering
    \includegraphics[width=0.75\columnwidth]{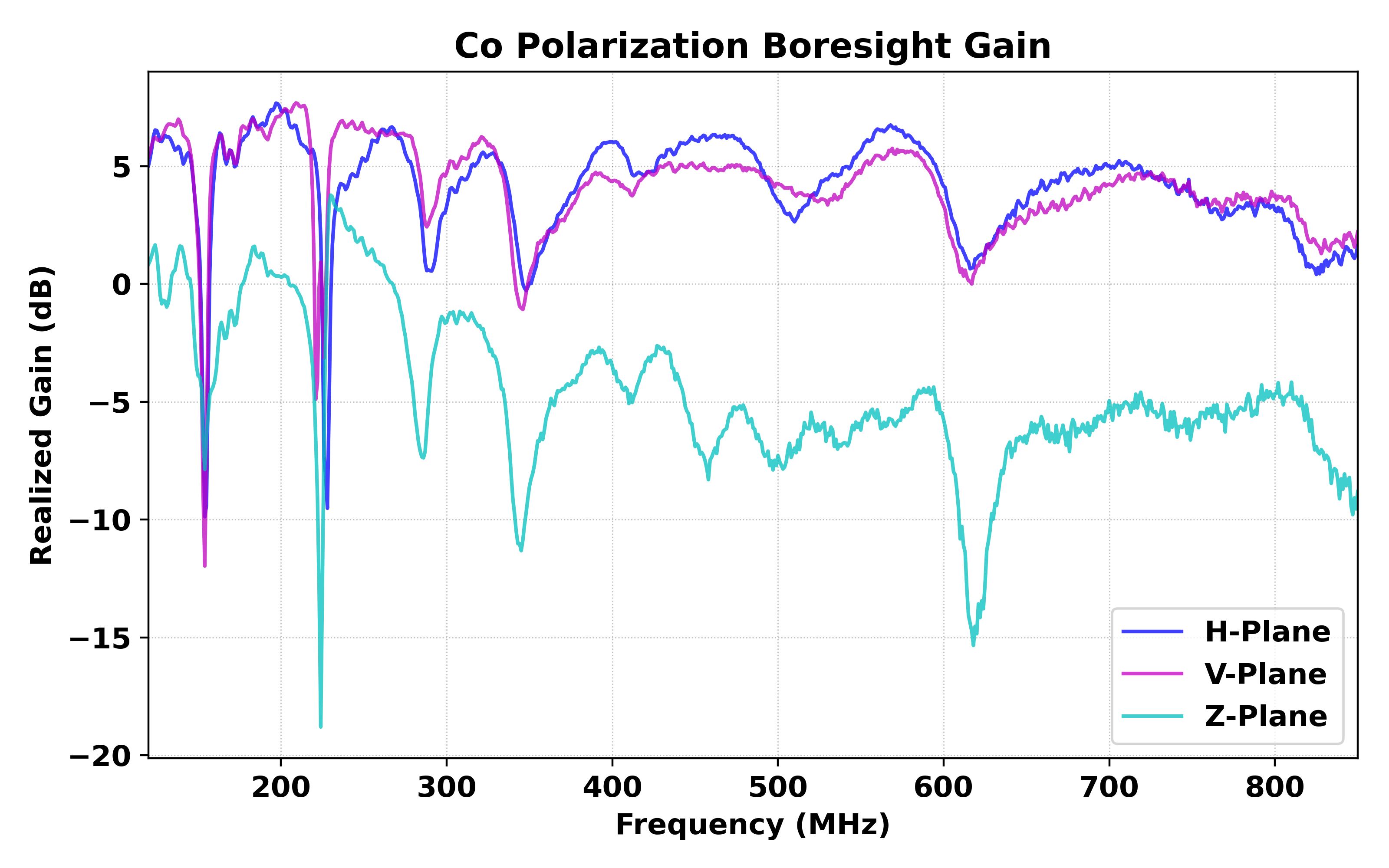}
    \includegraphics[width=0.75\columnwidth]{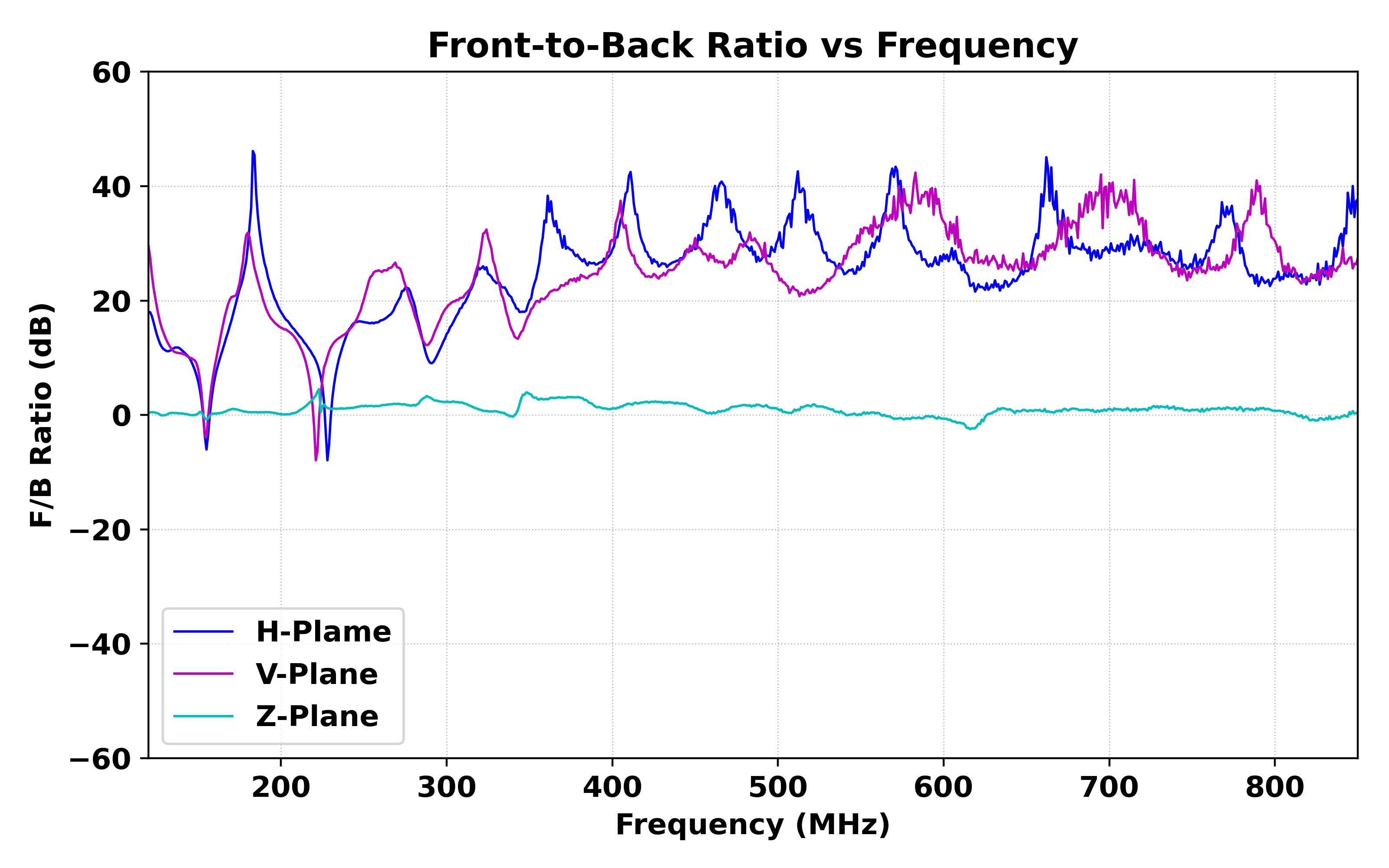}
    \caption{LPDA reference antenna gain metrics. \textbf{Top:} Co-polarization boresight gain as a function of frequency for the H-, V-, and Z-Planes. \textbf{Bottom:} Front-to-back (F/B) ratio versus frequency for the same Planes.}
    \label{fig:lpda_gain}
\end{figure}

For the LPDA reference antenna, Fig.~\ref{fig:lpda_gain} presents broadband gain characteristics in the H-, V-, and Z-Planes. The co-polarization boresight gain remains stable across frequency, while the front-to-back ratio demonstrates strong main-lobe directivity and effective suppression of backlobes in the H- and V-Planes. As expected, the Z-Plane front-to-back ratio centers near 0~dB, consistent with dipole-like radiation behavior.

\subsection{Validation with Simulation}

\begin{figure}[t]
    \centering
    \includegraphics[width=0.75\columnwidth]{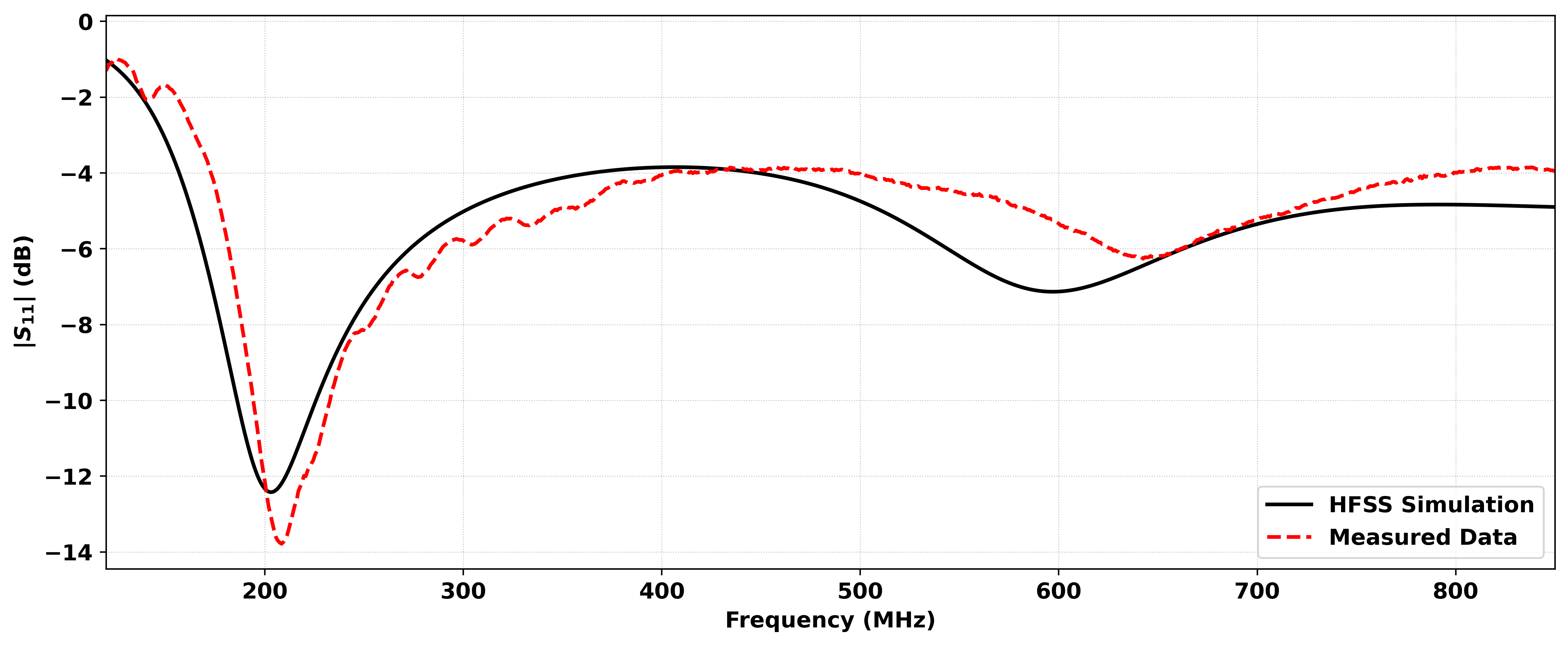}\\
    \includegraphics[width=0.75\columnwidth]{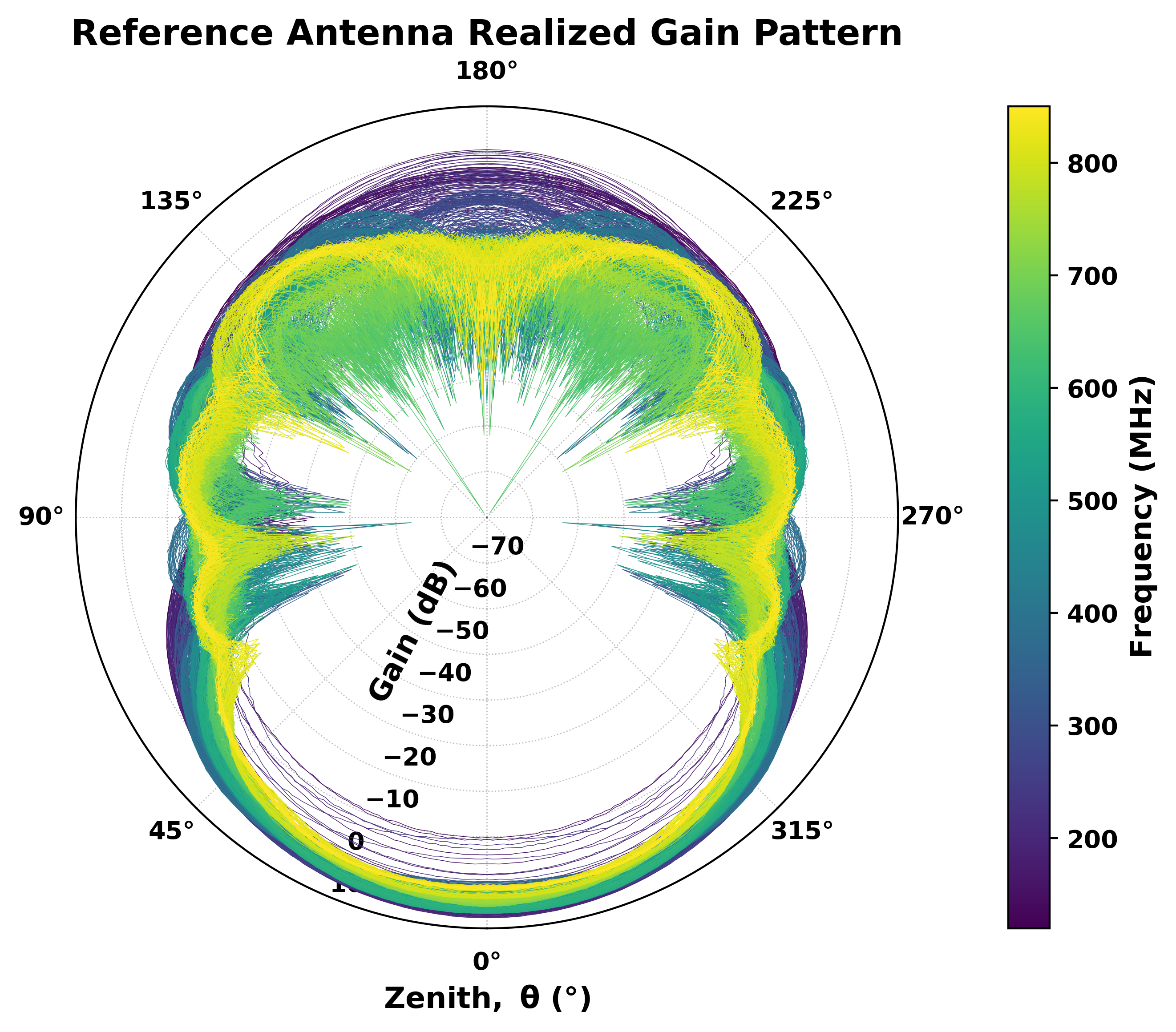}
    \includegraphics[width=0.75\columnwidth]{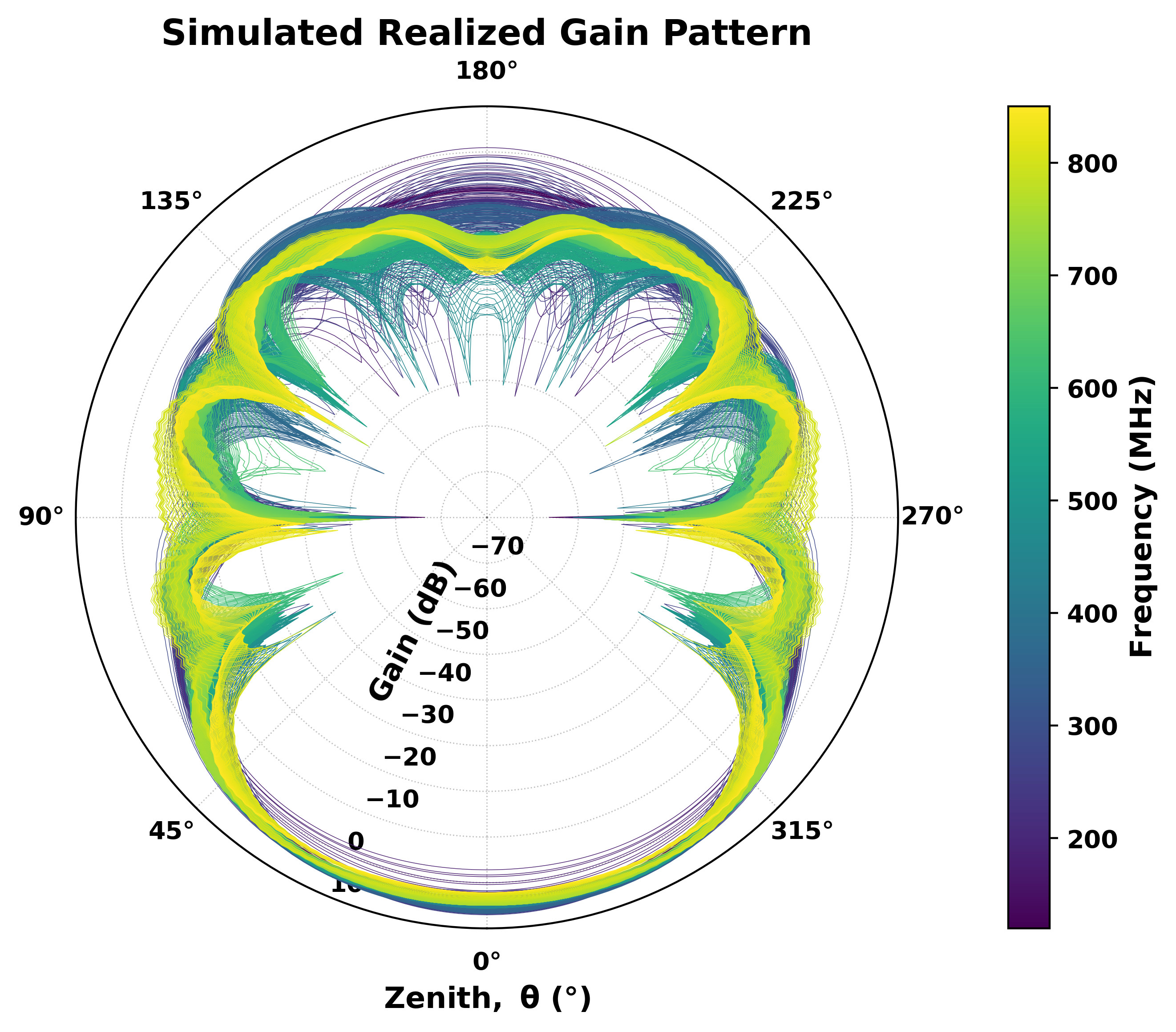}
    \caption{Validation of measurement and simulation. \textbf{Top:} Comparison of measured and HFSS-simulated $|S_{11}|$ for the ARA BVPol antenna. \textbf{Middle:} Measured LPDA realized gain pattern in the H-Plane. \textbf{Bottom:} Corresponding WIPL-D simulated realized gain pattern from NuRadioMC.}
    \label{fig:validation}
\end{figure}

To validate the measurement and analysis workflow, direct comparisons were performed between measured data and electromagnetic simulation results. Fig.~\ref{fig:validation} compares the measured $|S_{11}|$ response of the ARA BVPol antenna with HFSS simulations, showing good agreement in resonance frequency, bandwidth, and overall impedance behavior.

The lower panels of Fig.~\ref{fig:validation} compare measured realized gain patterns for the LPDA reference antenna with WIPL-D simulations generated within the NuRadioMC framework~\cite{NuRadioMC}. Close agreement is observed in main-lobe structure, angular dependence, and frequency evolution, supporting the accuracy of the calibration methodology and analysis pipeline. Minor discrepancies are attributed to fabrication tolerances, measurement uncertainties, and environmental effects.

\section{Conclusion}

This paper presented AAFIYA, a modular Python toolkit for automated antenna characterization using measurement and simulation data. The toolkit enables reproducible extraction of key antenna performance metrics, including S-parameters, impedance, realized gain, beam patterns, polarization metrics, and calibration-based yield estimation, within a unified and extensible analysis framework.

The capabilities of AAFIYA were demonstrated using measurements from an electromagnetic anechoic chamber involving Log Periodic Dipole Array reference antennas and Askaryan Radio Array Bottom Vertically Polarized antennas over a wide frequency range. The extracted impedance and gain characteristics show good agreement with full-wave electromagnetic simulations from HFSS and WIPL-D, validating both the measurement procedure and the analysis pipeline.

AAFIYA provides a flexible foundation for antenna research and instrumentation studies and can be readily adapted to new experimental configurations or datasets. Future developments will include integration with automated optimization and data-driven techniques, such as evolutionary design approaches demonstrated in recent antenna optimization studies~\cite{Rolla2023GENETIS}.


\section*{ACKNOWLEDGEMENT}
This research was supported in part by the National Science Foundation under Grants Nos. 2310096, 2310126, and 2012989, and by the Ralston Dissertation Fellowship from the Department of Physics and Astronomy, University of Kansas. The authors also acknowledge the Center for Remote Sensing of Ice Sheets (CReSIS) at the University of Kansas for access to the electromagnetic anechoic chamber used in the measurements. The authors further thank the Askaryan Radio Array (ARA) Collaboration for permission to test and characterize the ARA antennas.



%

\end{document}